\documentclass[11pt]{article}
\usepackage{latexsym,amsmath,amssymb}

\newtheorem{protocol}{Protocol}
\newtheorem{proposition}{Proposition}
\newtheorem{claim}{Claim}

\newcommand{\BF}[1]{\mathbf{#1}}   
\newcommand{\ket}[1]{|#1\rangle}                    

\newcommand{\Proof}{{\flushleft\it Proof.}}
\newcommand{\EndProof}{{$\Diamond$}}

\batchmode

\makeatletter
\def\@normalsize{\@setsize\normalsize{12pt}\xpt\@xpt
\abovedisplayskip 11pt plus2pt minus5pt
\belowdisplayskip \abovedisplayskip \abovedisplayshortskip \z@ plus3pt
\belowdisplayshortskip 6pt plus3pt minus3pt\let\@listi\@listI}
\def\subsize{\@setsize\subsize{12pt}\xipt\@xipt}
\def\section{\@startsection{section}{1}{\z@}{24pt plus 2 pt minus 2 pt}
{12pt plus 2pt minus 2pt}{\large\bf}}
\def\subsection{\@startsection {subsection}{2}{\z@}{12pt
plus 2pt minus 2pt}{12pt plus 2pt minus 2pt}{\subsize\bf}}

\makeatother
\begin{document}
\title{\bf Secure Communication Using Qubits\footnote{Submitted to Crypto 2005
on February 14, 2005}}
\author{Saied Hosseini-Khayat\footnote{University of Wollongong (Dubai Campus), UAE,
and Ferdowsi University, Iran, \it{saiedk@uow.edu.au}}
, Iman Marvian\footnote{Sharif University of Technology, Iran}
}
\date{}

\maketitle              

\begin{abstract}\it
A two-layer quantum protocol for secure transmission of data using
qubits is presented. The protocol is an improvement over the BB84 QKD protocol.
BB84, in conjunction with the one-time pad algorithm, has been shown to
be unconditionally secure. However it suffers from two drawbacks:
(1) Its security relies on the assumption that Alice's qubit
source is perfect in the sense that it does not inadvertently
emit multiple copies of the same qubit. A multi-qubit
emission attack can be launched if this assumption is violated.
(2) BB84 cannot transfer predetermined keys; the keys it can
distribute are generated in the process.
Our protocol does not have these drawbacks.

As in BB84, our protocol requires an
authenticated public channel so as to detect an intruder's interaction
with the quantum channel, but unlike in symmetric-key cryptography,
the confidentiality of transmitted data does not rely on a shared secret key.\\

{\bf Key words}: quantum cryptography, key distribution, data security.
\end{abstract}

\section{Introduction}\label{sec:intro}

Since the appearance of the BB84 protocol \cite{bb84},
a growing class of quantum
cryptographic protocols has emerged.
BB84 and its variants \cite{gisin2002,lomo2000} aim at
providing perfect security in transferring classical data
between two parties.
However, these protocols are key establishment protocols rather
than data transfer protocols.
They provide ultimate data security by generating a random
sequence of bit which is shared between Alice and Bob.
The sequence is then used as a one-time pad or as a
symmetric key. Because this sequence is generated in the process of protocol
execution, it cannot be known in advance. This can be a drawback in
applications where a predefined sequence of bits is to be distributed
securely. Our protocol can be used to avoid this drawback.
Also, though expensive, in some applications it may be
justifiable to use the quantum channel to send confidential data
as opposed to using the classical channel.
Our protocol can be used in those situations as well.

It has been shown that BB84 is unconditionally secure
\cite{biham2000,gottesman02}. However this fact depends on the assumption
that the qubit source does not emit multiple replicas of the same qubit.
In BB84, Alice generates a sequence of qubits which have been chosen at random
from one of two predefined bases.
She then sends a random sequence of bits using those qubits. Bob
makes measurement on the qubits upon receipt and then waits for the
basis information from Alice. The basis information comes through a
public authenticated channel. If the qubit source emits multiple replicas
for each intended qubit, an eavesdropper (Eve) on the quantum channel can
capture and preserve the qubits until the basis is announced by Alice.
At that time, Eve can perform measurement in the correct basis and obtain
the bits send by Alice. This is called a \emph{multi-qubit emission attack}.
Our protocol is not vulnerable to this kind of attack because information
crucial to correct measurement of the qubits is not transmitted or broadcast.

In our protocol, as in BB84 and other key exchange protocols,
the availability of an authenticated
(though not private) channel is an indispensable element.
This channel is the very essential means by which Bob can identify
Alice and differentiates her from Eve. In key exchange protocols
without an authenticated channel, Eve can launch a successful
man-in-the-middle attack. In our protocol, qubits are exchanged between
Alice and Bob, and they must make sure that the qubits are not altered or
inserted
by Eve. An authenticated channel is required for them to ensure this.

Our protocol has two major advantages over BB84 (explained above),
but it comes at a cost: The qubits must make a round trip
instead of a one-way trip. At a time when sending qubits over
long distances is a technical challenge, this requirement may sound too
troublesome.
However, there is a trade-off: A round trip must be weighted
against the challenge of making a single-qubit source.

This paper is organized as follows: In the next section, we
present the key ideas and properties used in our protocol.
Then in Section \ref{sec:prot}, the proposed protocols are presented.
In Section \ref{sec:anal}, we discuss the correctness and
security of our protocol.
At last in Section \ref{sec:con}, we summarize the results of
our paper. The appendix contains some of the technical points that
are necessary but are not crucial to understanding the concepts.


\section{Preliminaries}\label{sec:basics}

In this paper, we have borrowed some key concepts from BB84.
One key concept which is not from BB84 is the following:
Suppose Bob wants to send one bit to Alice.
Alice sends a random bit to Bob and she remembers this bit.
Bob performs an exclusive-or between Alice's bit and his bit
and sends the result to back Alice.
Alice can recover Bob's bit.
This is perfectly secure if
Eve cannot inspect Alice's bit on the way to Bob. If Eve inspects the bit
being sent from Bob to Alice, she will not know Bob's bit.
Of course, we should also prevent the possibility of a man-in-the-middle
attack by Eve.

It is hard to realize the above idea in classical cryptography.
However, in quantum cryptography this can be realized at least in
one way that is presented in this paper.

The whole idea of quantum cryptography revolves around the concept of
a \emph{qubit}. A qubit $\ket{\psi}$ is a quantum state vector in a two
dimensional Hilbert
space $\mathbb{H}_2$. A qubit can be measured with respect to any given basis
in $\mathbb{H}_2$. In quantum computation, the basis $\{\ket{0},\ket{1}\}$ is
called the \emph{computational basis}. Measuring a qubit
$\ket{\psi}=a\ket{0}+b\ket{1}$ in the computational basis will change it
to $\ket{0}$ or $\ket{1}$ with probabilities $|a|^2$ and $|b|^2$, respectively.
The observed outcome is logic
value '0' with probability $|a|^2$ or logic value '1' probability $|b|^2$.
Hereafter, whenever a qubit is "measured", we mean it is measured in the
computational basis.

In protocols discussed in this paper, Alice and Bob rotate qubits around a known
fixed axis in the Bloch sphere \cite[page 15]{nielsen2000}.
Without loss of generality and for the sake of convenience,
we choose this axis to be the $y$-axis.
To perform a rotation around the $y$-axis by an angle $\theta$, a given qubit
$\ket{\psi}$ must be operated upon by the following operator:
\[
R (\theta) = \left(
\begin{array}{cc}
  \cos{\frac{\theta}{2}}   &   -\sin{\frac{\theta}{2}} \\
  \sin{\frac{\theta}{2}}   &    \cos{\frac{\theta}{2}} \\
\end{array}
\right).
\]
It is easy to verify that for all angles
$\alpha$, $\beta$ and $\theta$, the following statements hold:
\[
R(\alpha) R(\beta) = R(\beta) R(\alpha) = R(\alpha+\beta),
\qquad R^{\dagger}(\theta) = R(-\theta).
\]
We will work with the following family of qubits:
\[
\ket{\psi(\theta)} = R(\theta) \ket{0} = \cos\frac{\theta}{2}\ket{0} +
\sin\frac{\theta}{2}\ket{1}.
\]
Since $\ket{0}$ is the state vector coinciding with the unit vector along the
$z$-axis in the Bloch sphere, then every $\ket{\psi(\theta)}$ is a
rotated version of the qubit $\ket{0}$ rotated by an angle $\theta$ around the
$y$-axis in the \emph{xz}-plane.
Also be reminded that any two qubits whose Bloch sphere representations are
colinear and point in opposite directions, e.g.\ $\ket{\psi(\theta)}$ and
$\ket{\psi(\theta+\pi)}$, are orthogonal to each other and hence they can be
detected and distinguished with perfect certainty.

Suppose $\ket{\psi(\theta)}$ is known by Alice to encode one of two logic values
0 or 1. If she makes measurement in the computational basis on this qubit, she
will obtain 0 with probability $\cos^2\frac{\theta}{2}$ and will obtain
1 with probability $\sin^2\frac{\theta}{2}$.
However if Alice knows the value of $\theta$, and if she operates
$R^{\dagger}(\theta)$ on this qubit before making measurement in the
computational basis, then she will obtain 0 with probability 1 (perfect
certainty).

Now suppose that Alice receives $\ket{\phi}$ and she knows this qubit
is either
$\ket{\psi(\theta)}$, or $\ket{\psi(\theta+\pi)}$.
Without knowing $\theta$, if she makes measurement, the outcome will
be 0 or 1 at random, the probability of each depending on the actual state
she received.
On the other hand, if Alice knows the value of $\theta$,
she can perform the unitary operation
$R^{\dagger}(\theta)\ket{\phi}$ and then followed by measurement in the
computational basis. With probability one, she will obtain 0 if
$\ket{\phi}=\ket{\psi(\theta)}$ or obtain 1 if
$\ket{\phi}=\ket{\psi(\theta+\pi)}.$
Therefore with known angle $\theta$, the states $\ket{\psi(\theta)}$ and
$\ket{\psi(\theta+\pi)}$ represent 0 and 1, respectively, and
can be distinguished with perfect certainty.

In what follows, we use the random variable $X$ to denote the binary information
transmitted, the random variable $Y$ to denote the binary information
received by an intended party, and the random variable $Z$ to denote the
information bit obtained by an intruder.
Adopting Shannon's definition, we deem a protocol as
unconditionally secure if
$H(X|Y)=0$ and $H(X|Z)=H(X)$. The first condition means that $Y$ reveals
everything about $X$.
The latter means that $Z$ reveals nothing about $X$; this is true if and only if
$X$ and $Z$ are independent.

We now prove some propositions that represent our key ideas and will be used
later. Consider a set of distinct angles
$\{\theta_0, \theta_1, \ldots, \theta_{n-1}\}$, all in $[0,2\pi].$
Suppose Alice prepares a qubit $\ket{\psi} = R(\theta_k+\pi X)\ket{0}$,
where $\theta_k$ was selected with probability $p_k$, and
$X \in \{0,1\}$ is a binary random variable (which is Alice's data).

\begin{proposition}\label{prop1}
If Alice sends the qubit to Bob. If Bob knows $\theta_k$, he
can recover $X$ without error.
\end{proposition}
\Proof\\
Bob can perform a unitary operation and obtain $\ket{\phi}$ as follows:
$\ket{\phi}=R^{\dagger}(\theta)\ket{\psi}$
He then makes measurement to obtain $Y$. We have:
\[
\ket{\phi} = \cos(\pi X/2) \ket{0} + \sin(\pi X/2) \ket{1}
\]
\[
\mbox{Prob}(Y=0 \mid X=0) = 1, \qquad
\mbox{Prob}(Y=1 \mid X=0) = 0
\]
\[
\mbox{Prob}(Y=0 \mid X=1) = 0, \qquad
\mbox{Prob}(Y=1 \mid X=1) = 1.
\]
This obviously implies: $H(X|Y)=0$. Hence the desired conclusion follows.
\EndProof

\begin{proposition}\label{prop2}
Alice sends the qubit to Bob. Eve, who does not knows
$\theta_k$, intercepts and rotates it by an arbitrary angle of her choice
$-\alpha$, makes measurement on the qubit and obtains a binary value $Z$.
There exist a probability distribution $p_k$ and a set
$\{\theta_0, \theta_1, \ldots, \theta_{n-1}\}$ such that
$H(X|Z)=H(X)$.
\end{proposition}
\Proof\\
We have:
\[
\ket{\psi} =
\cos[(\theta_k+\pi X)/2 ]\, \ket{0} +
\sin[(\theta_k+\pi X)/2 ]\, \ket{1}
\]
Eve's rotation produces:
\[
R(-\alpha) \ket{\psi} = \cos[(\theta_k+\pi X-\alpha)/2 ]\, \ket{0} +
\sin[(\theta_k+\pi X-\alpha)/2 ]\, \ket{1}.
\]
The probabilities for Eve's measurement outcome is:
\[
\mbox{Prob}(Z=0 \mid k,X=0) = \cos^2\frac{\theta_k-\alpha}{2}, \qquad
\mbox{Prob}(Z=0 \mid k,X=1) = \sin^2\frac{\theta_k-\alpha}{2},
\]
\[
\mbox{Prob}(Z=1 \mid k,X=0) = \sin^2\frac{\theta_k-\alpha}{2}, \qquad
\mbox{Prob}(Z=1 \mid k,X=1) = \cos^2\frac{\theta_k-\alpha}{2}.
\]
Therefore:
\[
\mbox{Prob}(Z=0 \mid X=0) = \mbox{Prob}(Z=1 \mid X=1) =
\sum^{n-1}_{k=0} p_k \cos^2\frac{\theta_k-\alpha}{2},
\]
\[
\mbox{Prob}(Z=0 \mid X=1) = \mbox{Prob}(Z=1 \mid X=0) =
\sum^{n-1}_{k=0} p_k \sin^2\frac{\theta_k-\alpha}{2}.
\]
The variables $X$ and $Z$ are independent if and only if:
\[ \mbox{Prob}(Z=0 \mid X=0) = \mbox{Prob}(Z=0 \mid X=1), \]
\[ \mbox{Prob}(Z=1 \mid X=0) = \mbox{Prob}(Z=1 \mid X=1). \]
Therefore $H(X|Z)=H(X)$ if and only if:
\begin{equation}\label{eq:33}
\sum^{n-1}_{k=0} p_k \cos^2\frac{\theta_k-\alpha}{2} =
\sum^{n-1}_{k=0} p_k \sin^2\frac{\theta_k-\alpha}{2}.
\end{equation}
We must find $p_k$'s and $\theta_k$'s such that
Equation (\ref{eq:33}) holds regardless of the choice of $\alpha$.
In Appendix A, Claim 1, it is shown that this is possible by setting
\[
n \ge 2, \quad p_k = \frac{1}{n}, \quad \theta_k = \frac{2k\pi}{n},
\quad \mbox{for } k =0,1,2,\ldots,n-1.
\]
Thus the conclusion follows. \EndProof

\begin{proposition}\label{prop3}
Suppose Alice sends the qubit to Bob.
Eve who does not knows $\theta_k$ intercepts the qubit and
performs a rotation by an angle of her choice $\alpha$. She then
makes measurement on
the qubit and then transmits the resulting qubit, denoted $\ket{\phi_1}$,
to Bob. If Bob knows $\theta_k$, and if he performs the following
unitary operation $\ket{\phi_2} = R^{\dagger}(\theta)\ket{\phi_1}$ followed
by a measurement of $\ket{\phi_2}$, then the channel between Alice and Bob
has an error probability equal to:
\[
\mbox{Prob}(\mbox{error}) =
\sum_{k=0}^{n-1}  p_k \left[
\sin^2\frac{\theta_k}{2} \, \cos^2\frac{\theta_k-\alpha}{2} +
\cos^2\frac{\theta_k}{2} \, \sin^2\frac{\theta_k-\alpha}{2} \right].
\]
\end{proposition}
\Proof\\
We are looking for $P(Y|X)$ and do this by conditioning on $Z$ and $k$.
\[
P(Y|X) = \sum_k \sum_Z P(Y|X,Z,k)\, P(Z|X,k)\, P(k|X).
\]
Since $k$ is chosen with probability $p_k$ independent of $X$, then
\[
P(k|X) = p_k, \quad k=0,1,\ldots,n-1.
\]
After Eve's measurement, there are two cases to consider:
\begin{enumerate}
\item $Z=0$. In this case: $\ket{\phi_1}=\ket{0}$. After Bob's rotation. we have
\[
\ket{\phi_2}= R^{\dagger}(\theta) \ket{\phi_1} =
\cos\frac{\theta_k}{2}\,\ket{0} - \sin\frac{\theta_k}{2}\,\ket{1}.
\]
After Bob's measurement we have:
\[
\mbox{Prob}(Y=0 \mid k,X,Z=0) = \cos^2\frac{\theta_k}{2}, \qquad
\mbox{Prob}(Y=1 \mid k,X,Z=0) = \sin^2\frac{\theta_k}{2}.
\]
\item $Z=1$. In this case: $\ket{\phi_1}=\ket{1}$. After Bob's rotation. we have
\[
\ket{\phi_2}= R^{\dagger}(\theta) \ket{\phi_1} =
\sin\frac{\theta_k}{2}\,\ket{0} + \cos\frac{\theta_k}{2}\,\ket{1}.
\]
After Bob's measurement we have:
\[
\mbox{Prob}(Y=0 \mid k,X,Z=1) = \sin^2\frac{\theta_k}{2}, \qquad
\mbox{Prob}(Y=1 \mid k,X,Z=1) = \cos^2\frac{\theta_k}{2}.
\]
\end{enumerate}
The conditional probabilities for $Z$ given $X,k$ were obtained in
Proposition (\ref{prop2}). Using those, we can compute the following:
\[
\mbox{Prob}(Y=0 \mid X=0) = \mbox{Prob}(Y=1 \mid X=1) =
\sum_{k=0}^{n-1} p_k \left[
\sin^2\frac{\theta_k}{2} \, \sin^2\frac{\theta_k-\alpha}{2} +
\cos^2\frac{\theta_k}{2} \, \cos^2\frac{\theta_k-\alpha}{2} \right],
\]
\[
\mbox{Prob}(Y=0 \mid X=1) = \mbox{Prob}(Y=1 \mid X=0) =
\sum_{k=0}^{n-1}  p_k \left[
\sin^2\frac{\theta_k}{2} \, \cos^2\frac{\theta_k-\alpha}{2} +
\cos^2\frac{\theta_k}{2} \, \sin^2\frac{\theta_k-\alpha}{2} \right].
\]
Note that:
\[
\mbox{Prob}(\mbox{error}) = P(X=0)P(Y=1|X=0)+P(X=1)P(Y=0|X=1)
\]
Therefore regardless of $P(X)$ we find:
\begin{equation}\label{eq:71}
\mbox{Prob}(\mbox{error}) =
\sum_{k=0}^{n-1}  p_k \left[
\sin^2\frac{\theta_k}{2} \, \cos^2\frac{\theta_k-\alpha}{2} +
\cos^2\frac{\theta_k}{2} \, \sin^2\frac{\theta_k-\alpha}{2} \right].
\end{equation}
The proof is complete.
\EndProof\\

\begin{proposition}\label{prop4}
In Proposition \ref{prop3}, suppose we set
\[
n > 2, \quad p_k = \frac{1}{n}, \quad \theta_k = \frac{2k\pi}{n},
\]
then
\[
\mbox{Prob}(\mbox{error}) \ge \frac{1}{4},
\]
with equality when $\alpha=0$.
\end{proposition}
\Proof \; See Appendix A, Claim 2. \EndProof\\

This proposition means that by certain choice of $p_k$'s and $\theta_k$'s
(as prescribed above) we can be sure that Eve's interaction will be detected
with probability at least 1/4.
Also, from the point of view of Eve, who wants to minimize her
probability of being detected, this means that her best strategy is
to do measurement only without any rotations.
This result will be used in the analysis
of our qubit authentication protocol (Protocol 1).

\section{Proposed Protocol}\label{sec:prot}

First we present a protocol for sending authenticated qubits. Then we present a
protocol for sending secure data using authenticated qubits.

\subsection{Sending Authentic Qubits}\label{sec:saq}

We define an \emph{authentic qubit} as a qubit which has been sent by Alice
and received by Bob such that Eve has not altered it in any way along the way.
Suppose Alice wants to send authentic qubits to Bob. For convenience, we
restrict our discussion to the family of qubits residing in the \emph{xz}-plane.

\begin{protocol}\rm
Suppose Alice has a sequence of $N$ qubits
$ \ket{\psi_1}, \ket{\psi_2}, \ldots, \ket{\psi_N}, $
and wishes to send them to Bob in an authenticated way. We
assume that a classical bidirectional authenticated channel exists between
Alice and Bob. Also assumed is a set of publicly-known
distinct angles $ \theta_k = \frac{2k\pi}{n},
k=0,1,\ldots,n-1.$
\begin{enumerate}
\item Alice selects $M$ integers uniformly at random in $\{0,1,\ldots,n-1\}$,
denoted $k_1, k_2, \ldots, k_M$.
\item Alice creates qubits $ \ket{\phi_1}, \ket{\phi_1}, \ldots, \ket{\phi_M}, $
such that for each $i \in \{1,2,\ldots,M\}$
\[
\ket{\phi_i}= R(2k_i\pi/n)\,\ket{0}.
\]
We call these the \emph{check qubits}.
\item Alice generates $M$ distinct random integers $j_1, j_2, \ldots, j_M$ in
$\{1,2,\ldots,N+M\}$. She then creates a frame of length $N+M$ and inserts
each $\ket{\phi_i}$ at location $j_i$ in the frame.
Then the sequence of qubits
$ \ket{\psi_1}, \ket{\psi_2}, \ldots, \ket{\psi_N}, $
are inserted at the empty locations in the frame preserving their order.
\item Alice sends the frame to Bob. Bob receives the frame.
\item Alice sends the following data to Bob:
\[ (j_1,k_1),(j_2,k_2),\ldots,(j_M,k_M). \]
These data are sent via the classical authenticated channel.
Eve can read these data.
\item Knowing $j_i$'s, Bob extracts the sequence of
$\ket{\phi_i}$'s from the frame, and for each $i=1,2,\ldots,M$, he performs
$R^\dagger(2k_i\pi/n) \ket{\phi_i}$. Then he
measures the resulting state. The outcome must be a logical '0' for
all $i \in \{1,2,\ldots,M\}$.
If this condition does not hold for any $\ket{\phi_i}$, the frame
is said to have an \emph{authentication error}.
\item If there is an authentication error, Bob notifies Alice and both
drop the frame. Otherwise, the frame is considered unaltered and
the sequence of $\ket{\psi_i}$'s is deemed to be authentic.
\end{enumerate}
\end{protocol}

At a given level of certainty (determined by parameters $N$ and $M$),
this protocol uses a classical authenticated channel to create an authenticated
quantum channel. This protocol is analyzed in Section \ref{sec:anal}.
We use this protocol as a tool to prevent the man-in-the-middle attack
in our confidential data transfer protocol (Protocol 2).

\subsection{Sending Confidential Qubits}\label{sec:ssq}

In this section, we present a protocol that enables sending confidential
data over the quantum channel. In this protocol, qubits make a round trip
from Alice to Bob to Alice and undergo a unitary operation by Bob along the way.
The protocol is described below.

\begin{protocol}\rm
Suppose Bob has a sequence of $N$ data bits $x_1,x_2,\ldots,x_N$ where
$x_i \in \{0,1\}$.
The existence of a classical bidirectional authenticated channel exists
between Alice and Bob is assumed.
\begin{enumerate}
\item Alice generates a sequence of random integers $k_1,k_2,\ldots,k_N$
where $k_i \in \{0,1,\ldots,n-1\}$. She keeps these integers confidential
to herself.
\item Alice creates $n$ qubits
$ \ket{\psi_{11}}, \ket{\psi_{12}}, \ldots, \ket{\psi_{1N}}, $
such that $\ket{\psi_{1i}} = R(\frac{2 \pi k_i}{n})\ket{0}.$
\item Alice sends the $N$ qubits to Bob using Protocol 1.
\item Bob receives the $N$ authenticated qubits from Alice.
On each $\ket{\psi_{1i}}$, without making measurement, he performs the unitary
operation $R(\pi x_i)$ to produce the qubit sequence
\[ \ket{\psi_{21}}, \ket{\psi_{22}}, \ldots, \ket{\psi_{2N}} \]
where $\ket{\psi_{2i}} = R(\pi x_i) \ket{\psi_{1i}}. $
This sequence is sent to Alice in an authenticated way using Protocol 1.
\item Alice receives the $N$  authenticated qubits.
On each qubit $\ket{\psi_{2i}}$, she perform unitary operation
$R^\dagger(\frac{2\pi k_i}{n}).$ The resulting sequence denoted
\[ \ket{\psi_{31}}, \ket{\psi_{32}}, \ldots, \ket{\psi_{3N}} \]
is then measured to produce bit sequence $y_1,y_2,\ldots,y_N$.
This sequence is deemed to be the sequence sent by Bob.
\end{enumerate}
\end{protocol}

As shown in Propositions 1 through 4, there seems to be no advantage
in selecting $n>3$. Therefore we propose setting $n=3$ in both protocols.

\section{Analysis}\label{sec:anal}

In this section, we analyze the two protocols to show that they indeed
serve their intended purposes. In our analysis, we take advantage of
the propositions set forward in the preliminary section.
Since our primary purpose in this paper is introducing a protocol for
confidential data transfer, We analyze Protocol 1 first while assuming Protocol
2 serves its purpose perfectly (i.e., sending authentic qubits).

\subsection{Analysis of Protocol 2}\label{sec:anal2}

We analyze this protocol in two steps: First we show it is correct,
second, we show it is secure.

\subsubsection*{Correctness}
We show that Protocol 2 works correctly by invoking Proposition 1.  Notice
that Alice prepares each qubit $\ket{\psi_{1i}}$
at a certain angle $\frac{2 \pi k_i}{n}$ which she only knows.
Bob encodes his data bit in the qubit by performing
$R(\pi x_i) \ket{\psi_{1i}}$. Doing so means that Alice's qubit either
remains the same or is rotated by an angle $\pi$. In Proposition 1, Alice
encodes her data and sends to Bob. If Bob knows the original qubit's angle
can recover Alice's data. In the case of Protocol 2 (which is slightly
different), it is Bob who encodes his data, but since he does not know the
qubit's angle, he sends it to Alice
who (by the reason of Proposition 1) can recover Bob's data without error.
Therefore, Protocol 2 is correct.

\subsubsection*{Security}
We show that Protocol 2 is secure.
This means that an intruder (Eve)
cannot recover Bob's data. We do this in the following:

\begin{itemize}
\item[a)] Note that if Eve can find out the angle
$\frac{2 \pi k_i}{n}$ for each qubit $\ket{\psi_{1i}}$, she, like Alice, can
recover Bob's data.
An invocation of Proposition 2 will prove that this is not possible.

First notice that Protocol 2 uses the specific set of $p_k$'s and $\theta_k$'s
that satisfies Proposition 2 for security. (Thus Equation (\ref{eq:33})
is satisfied.) Next, notice that the first trip of a qubit from Alice to Bob
corresponds to the same trip in Proposition 2 with the exception that Alice's
data is set to a constant logic '0'.
Since the condition for Equation (\ref{eq:33}) is met by our specific choice
of $p_k$'s and $\theta_k$'s, therefore:
\[
\mbox{Prob}(Z=0 \mid X=0) = \mbox{Prob}(Z=1 \mid X=0) = \frac{1}{2}.
\]
This means that when Eve (intercepts and ) measures each $\ket{\psi_{1i}}$ (
possibly after a rotation of her choice), then she observes logic '0' or '1'
with equal probability, regardless of Alice's choice of angle
$\frac{2 \pi k_i}{n}$. Therefore, she can't get any information about the
qubit's angle
in its trip from Alice to Bob.

\item[b)] Although the conclusion in item b) is good news, note that if there
are only two angles for Alice to choose from (i.e., $n=2$) then that conclusion
will break down because Eve know Alice's data is '0', and since Alice can only
prepare one of the two states $R(0) \ket{0}$ or $R(\pi)\ket{0}$, then
Eve can find out the angle with certainty because the two states are orthogonal.
Therefore we have to impose the condition $n>2$ for the sake of security.

\item[c)] To show that performing Protocol 1 is essential in the first leg
of the trip, we assume that Alice sends the qubits to Bob without performing
Protocol 1 (thus modifying Step 3 in Protocol 2).
While Eve can get no information about the specific choice of
qubit angles by Alice, Eve is still able to launch the following
man-in-the-middle attack:\\
Eve intercepts and performs measurement on Alice's qubits.
This allows her to know the
state of each qubit (they will be either $\ket{0}$ or $\ket{1}$.)
She then sends these to Bob. (Or she might as well drop Alice's qubits and send
her own.) Bob, who is not aware, encodes his data in Eve's qubits and send them
to Alice. Eve intercepts, makes measurement (thus perfectly recovering Bob's
data), and then sends arbitrary qubits to Alice. Alice's measurements (after
appropriate rotations that she knows) will produce gibberish. But Eve has
managed to steal Bob's data.
\emph{Thus} it is absolutely essential that Bob knows that the qubits he
receives are authentic; that is, they come from Alice without Eve having the
opportunity to intervene.
This is the job of Protocol 1 (assuming it works flawlessly.)

\item[d)] On the return trip for qubits (from Bob to Alice), Eve can intercept
and make measurement on the qubits. These qubits carry data, however a direct
invocation of Proposition 2 shows that no data can be gained by Eve if she
intercepts (Apply Proposition 2).
However, if Bob does not use Protocol 1 (in Step 4)
in transmitting the qubits back to Alice,
a malicious Eve can intercept, drop the qubits and insert her own.
In this case, Alice will receive gibberish instead of meaningful data.
This is a disruption of communication as opposed to compromised security.

\item[e)] In this and the next items, we show security against multi-qubit
emission attack. Suppose a qubit
source used by Alice is imperfect and for each intended qubit it creates
multiple replicas in exactly the same state as the original.
This can create a severe security problem in BB84: Eve can capture and preserve
a replica qubit and wait until the angle information is announced on the
public channel. She then can perform measurement in the correct basis and
recover the data.
This cannot happen in Protocol 2 because no such information concerning the
data carrying qubits is ever sent.  Only in Protocol 1, angle information about
the check qubits are sent on the public channel. This cannot give away
information about Bob's data. We will deal with the check qubits when we
analyze Protocol 1.

\item[f)] Another attack using multiple qubits is the following:
In Step 3, Protocol 2, Alice sends qubits to Bob. Because of the multiple-qubit
imperfection, Eve can capture and preserve a qubit replica for each original
qubit Alice sends. During the execution of protocol 1, the check qubits are
identified,
therefore Eve can throw them away. She keeps the non-check qubits which
are intended for carrying data. In the return trip, Bob sends the data carrying
qubits encoded with his data.  Not knowing the qubit's angles,
Eve cannot extract any information from the
replicas she received from Alice nor from the originals or replicas
she receives from Bob. (Now assume Protocol 1 is not perform in the
return trip at Step 4.)
Eve can drop Bob's qubits and insert her own. These are
the replicas she kept now encoded with her fake data.
This data can be received and understood by Alice. This attack affects
the integrity of data rather than its confidentiality.

However, this attack cannot succeed because of using Protocol 1 at Step 4.
At Step 4, Bob sends his qubits ($\ket{\psi_i}$'s) along with the check
qubits ($\ket{\phi_i}$'s) he inserts
in the frame. Eve who receives all these qubits has no information about
the position of these qubits. Therefore she cannot
replace the non-check qubits with her own. She can at best make a random
guess about the position of $N$ non-check qubits and attempt to replace them.
Eve must be at least 1 in ${{N+M}\choose N}$ lucky to succeed.
This means that by a choice of sufficiently large $N$ and $M$, her
luck can be made completely insignificant.
Remember that these two parameters can be set to exceed any given numbers.
If necessary, an appropriate padding algorithm can extend the length $N$
of Bob's data to a required minimum.
Therefore, we have justified the use of Protocol 1 in Step 4, and thus
removed the possibility of using multiple-qubit attack.

\end{itemize}

Up to now, we have discussed the various ways Protocol 2 can be attacked,
and have shown that none can succeed. As it was seen, the security of
Protocol 1 was a key assumption. In the next section, we turn to analyzing
this assumption.

\subsection{Analysis of Protocol 1}\label{sec:anal1}

We show that Protocol 1 allows Alice to send Bob authenticated qubits.
Authenticity of these qubits are essential for guaranteeing the security
of Protocol 2. It is easy to see that Protocol 1 works correctly if Eve
is not present.

Suppose Alice sends to Bob a frame of $N+M$ qubits using Protocol 1. The idea
is that Eve cannot interact with the qubits with vanishing probability of
being detected by Bob.
According to Propositions 3 and 4, because of the special choice of
$p_k$'s and $\theta_k$'s, if Eve performs a measurement (possibly after an
arbitrary rotation) on any one of the check qubits, the probability
that her action produces an error detectable by Bob is at least $P_e = 1/4$.

Now suppose Eve decides to inspect $L$ qubits at random in the frame.
The probability that out of $L$ qubits, $k$ are check qubits is as follows:
\[
\mbox{Prob}(\mbox{Eve picks $k$ check qubits}) =
\frac{ {N \choose {L-k}} {M \choose k }  }{ {{M+N} \choose L }}.
\]
The probability that Eve goes undetected given she inspects $k$ check qubits is
as follows:
\[
\mbox{Prob}(\mbox{Eve undetected} \mid k) = (1-P_e)^k.
\]
Therefore the probability that Eve's interaction is detected when she inspects
$L$ qubits in the frame at random is:
\begin{equation}\label{eq:13}
\mbox{Prob}(\mbox{Eve undetected when she inspects $L$ qubits}) = \sum_{k=0}^L
(1-P_e)^k \,
\frac{ {N \choose {L-k}} {M \choose k }  }{ {{M+N} \choose L }}.
\end{equation}
It is easy to compute, either by using
Equation (\ref{eq:13}) or directly, the following extreme cases:
\begin{itemize}
\item[I.] Eve decides to inspect all qubits in a frame.
The probability that her action is not detected is:
\[
\mbox{Prob}(\mbox{Eve undetected}) = (1-P_e)^M.
\]
If she succeeds, she has been able to measure all $N$ qubits that were
supposed to be protected. This event can be made arbitrarily improbable
by increasing the number of check qubits $M$.
\item[II.] Eve decides to inspect only one qubit in a frame.
The probability that her action is not detected:
\[
\mbox{Prob}(\mbox{Eve undetected}) = 1 - \frac{M P_e}{N+M}.
\]
If she succeeds, at best she has been able to measure only one of the $N$
qubits.
The probability of this event can only be made as small as $(1-P_e)$ by
increasing $M$.
\end{itemize}
In both cases, increasing $M$ implies decreasing the probability that
Eve is not detected.

In another attack on this protocol, Eve does not measure the qubits but
attempts to replace them with her own.
Suppose she wishes to send $L$ qubits of her own.
She must choose $L<N$ qubits in the frame to be replaced.
Lack of any knowledge about the position of check qubits makes her
guess randomly. She will be lucky if she selects none of the check qubits.
The probability of this event is
$
{ {N \choose L} } / { {{M+N} \choose L }}.
$
This probability can be made a small as desired by choosing a sufficiently large
number $M$ of check qubits.
In general, by increasing parameter $M$, we can make the probability that Eve's
measurement or replacement of $L$ qubits goes undetected as small as desired.

\section{Conclusion}\label{sec:con}

We introduced and analyzed two novel quantum protocols that together
allow a secure transfer of classical data bits.
The first protocol enables Alice and Bob to exchange authenticated qubits.
The second protocol, which makes use of the first one, enables Alice and Bob
to exchange data bits securely.
The combination has the following merits:
\begin{itemize}
    \item[a)] It can send data over the quantum channel securely.
    This is especially useful in key distribution applications where keys are
    generated in advance rather than on the fly.
    BB84 and its variants cannot transfer predetermined keys.
    \item[b)] In BB84, some information leaks out, therefore a process called
    privacy amplification is required by that protocol.
    In our protocol, no information about the data leaks out.
    \item[c)] Our protocol, unlike BB84, is not vulnerable to multi-qubit
    emission attack.
\end{itemize}
One disadvantage of our protocol is its round trip requirement. This is a price
to pay to protect against multi-qubit emission attack.
While we have discussed most probable attack scenarios, there is
more to be done to prove the unconditional security of this protocol.

\section*{Appendix A}\label{appx:1}
\begin{claim}\rm
We show that for all $\alpha$, if
\[
n \ge 2, \quad p_k = \frac{1}{n}, \quad \theta_k = \frac{2k\pi}{n},
\quad \mbox{for } k =0,1,2,\ldots,n-1.
\]
then the following holds:
\begin{equation}\label{eq:43}
\sum^{n-1}_{k=0} p_k \cos^2\frac{\theta_k-\alpha}{2} =
\sum^{n-1}_{k=0} p_k \sin^2\frac{\theta_k-\alpha}{2}.
\end{equation}
\end{claim}
\Proof \; Equation (\ref{eq:43}) implies:
\[
\sum^{n-1}_{k=0} p_k \cos(\theta_k-\alpha) = 0.
\]
Note that the left hand side of the above equation can be written as:
\[
\sum^{n-1}_{k=0} p_k \, \BF{r}_{\theta_k} \cdot \BF{r}_{\alpha} = 0,
\]
where $\BF{r}_{\theta_k}$ and $\BF{r}_{\alpha}$ are two units vectors
located at the origin in the $xz$-plane making an angle $\theta_k$ and $\alpha$
with the $z$-axis, respectively.
The above equation must hold for any $\BF{r}_{\alpha}$, therefore we must have:
\[
\sum^{n-1}_{k=0} p_k \, \BF{r}_{\theta_k} = 0,
\]
If we set $ p_k = \frac{1}{n} $ then we must have:
\[
\sum^{n-1}_{k=0} \BF{r}_{\theta_k} = 0,
\]
This can be made true by setting:
\[
\theta_k = \frac{2k\pi}{n},
\quad \mbox{for } k =0,1,2,\ldots,n-1.
\]
The claim is proved. \EndProof

\begin{claim}\rm
Proposition \ref{prop4} is true.
\end{claim}
\Proof \;
The error probability given in Equation(\ref{eq:71}) can be re-written
using trigonometric identities as:
\[
\mbox{Prob}(\mbox{error}) =
\frac{1}{2} - \frac{1}{2} \sum_{k=0}^{n-1}  p_k \left[
    \cos^2\theta_k \, \cos\alpha +
\cos\theta_k \, \sin\theta_k \sin\alpha \right].
\]
By setting $p_k = \frac{1}{n}$, and $ \theta_k = \frac{2k\pi}{n},$
we arrive at:
\[
\mbox{Prob}(\mbox{error}) =
\frac{1}{2} - \frac{1}{2n} \sum_{k=0}^{n-1}
    \cos^2\frac{2k\pi}{n} \, \cos\alpha
    - \frac{1}{2n} \sum_{k=0}^{n-1}
\cos\frac{2k\pi}{n} \, \sin\frac{2k\pi}{n} \sin\alpha .
\]
Now note that for any $n>2$:
\[
\sum_{k=0}^{n-1} \cos\frac{2k\pi}{n} \, \sin\frac{2k\pi}{n} = 0.
\]
This can be easily verified by noticing that for $i>0$ the
terms corresponding to $k=i$ and $k=n-i$ cancel each other out.
For $k=0$, the term is already zero, and when $n$ is even, the
term corresponding to $k=n/2$ is also zero.

Other the other hand, notice that for any $n>1$:
\[
\sum_{k=0}^{n-1} \cos^2\frac{2k\pi}{n} = \frac{n}{2}.
\]
This is true because:
\[
\sum_{k=0}^{n-1} \cos^2\frac{2k\pi}{n} =
\frac{1}{2} \sum_{k=0}^{n-1} (1-\cos\frac{4k\pi}{n}) =
\frac{n}{2} - \frac{1}{2} \sum_{k=0}^{n-1} \cos\frac{4k\pi}{n}.
\]
(It can be easily shown that $\sum_{k=0}^{n-1} \cos\frac{4k\pi}{n}=0$.)
Therefore, finally:
\[
\mbox{Prob}(\mbox{error}) =
\frac{1}{2} - \frac{1}{4} \cos\alpha \ge \frac{1}{4},
\]
with equality when $\alpha=0$. \EndProof

%
%

\end{document}